\documentclass[aps,prl,twocolumn,amsfonts,superscriptaddress,amssymb,amsmath]{revtex4-1}

\usepackage{graphicx}
\usepackage[separate-uncertainty=true]{siunitx}
\usepackage{subcaption}
\usepackage{amssymb}
\usepackage{enumitem}
\usepackage{bbold}
\usepackage{color}
\newcommand{\ymo}{YMnO$_\mathrm{3}$}
\newcommand{\gs}{$P6^\prime_3cm^\prime$}
\newcommand{\es}{$P6^\prime_3c^\prime m$}
\usepackage{wasysym}

\begin{document}

\title{Tracking the ultrafast motion of an antiferromagnetic order parameter}

\author{Christian Tzschaschel}
\email{christian.tzschaschel@mat.ethz.ch}
\affiliation{Department of Materials, ETH Zurich, 8093 Zurich, Switzerland}

\author{Takuya Satoh}
\affiliation{Department of Physics, Kyushu University, Fukuoka 819-0395, Japan}

\author{Manfred Fiebig}
\affiliation{Department of Materials, ETH Zurich, 8093 Zurich, Switzerland}

\date{\today}

\begin{abstract}
The unique functionalities of antiferromagnets offer promising routes to advance information technology  \cite{Jungwirth16,Gomonay18}. Their compensated magnetic order leads to spin resonances in the THz-regime  \cite{Nemec18}, which suggest the possibility to coherently control antiferromagnetic (AFM) devices orders of magnitude faster than traditional electronics. However, the required time resolution, complex sublattice interations and the relative inaccessibility of the AFM order parameter pose serious challenges to studying AFM spin dynamics. Here, we reveal the temporal evolution of an AFM order parameter directly in the time domain. We modulate the AFM order in hexagonal \ymo\ by coherent magnon excitation and track the ensuing motion of the AFM order parameter using time-resolved optical second-harmonic generation (SHG). The dynamic symmetry reduction by the moving order parameter allows us to separate electron dynamics from spin dynamics. As transient symmetry reductions are common to coherent excitations, we have a general tool for tracking the ultrafast motion of an AFM order parameter.
\end{abstract}

\maketitle

Recently, the appealing properties of antiferromagnets culminated in the demonstration of both optically  \cite{Kimel09} and electrically  \cite{Wadley16} induced switching of the AFM order. This illustrates the general usefulness of antiferromagnets for future applications, yet the dynamics occurring during the switching process remain uncertain as time-resolved access to the AFM order parameter is notoriously difficult. Consequently, detecting its motion in the time-domain often relies on measuring one spin component, such as a transiently occuring uncompensated magnetic moment, and extrapolating the full order-parameter dynamics from models  \cite{Kimel09}. A detailed study of all spin components during a ferromagnetic spin precession, however, revealed previously unexpected insights  \cite{Acremann00}. Thus, a dependable analysis of the non-trivial spin dynamics during switching events in antiferromagnets necessitates direct and time-resolved experimental access to all the order-parameter components.

While the uncompensated component is optically accessible by the Faraday effect, detection of the compensated part of the order parameter usually depends on more indirect processes, like magnetic linear birefringence, which is quadratic in the sublattice magnetisation. This leads to a loss of directionality and, thereby, of the ability to distinguish between spin-reversed domain states  \cite{Wadley16, Saidl17, Tzschaschel17}. Here SHG, i.e.\ frequency doubling of a light wave in a material, is a powerful, symmetry-sensitive technique that can couple linearly to the AFM order parameter, thus maintaining its directional information  \cite{Sa2000, Fiebig05}. The leading SHG contribution is given by $P_i(2\omega) = \epsilon_0\mathcal{X}_{ijk}E_j(\omega)E_k(\omega)$, where $\epsilon_0$ is the vacuum permittivity, $\mathbf{E}$ is the electric field of the incoming light wave, $\mathcal{X}$ is the second-order susceptibility tensor and $\mathbf{P}$ is the induced polarisation oscillating at twice the frequency $\omega$ of the incident light  \cite{Shen84}. Owing to the direct coupling of $\mathcal{X}$ to the AFM order parameter, SHG is ideal for the static characterisation of antiferromagnets  \cite{Fiebig00b,Fiebig05,Chauleau17}. Despite this and despite the instantaneous nature of the SHG process, which allows probing ultrafast processes, time-resolved (TR) SHG studies of optically induced dynamics are scarce. The methodical difficulty to discriminate between genuine magnetic-order-parameter dynamics, affecting $\mathcal{X}_{ijk}$, and electron dynamics, affecting the linear optical properties at $\omega$ and $2\omega$, hampers the interpretation of TR-SHG measurements  \cite{Huber15}. In fact, distinguishing between order-parameter and electron dynamics is commonly believed to be a ``practical impossibility''  \cite{Kirilyuk10}. This is particularly true for thermal excitations, where redistribution of resonantly excited charge carriers leads to substantial changes in the linear optical properties  \cite{Satoh07,Matsubara09,Sala16,Sheu16}. 

An optical excitation, however, can also act as a non-thermal stimulus to the magnetic order. A well-known example is the inverse Faraday effect (IFE)  \cite{Kirilyuk10}, where a circularly polarised laser pulse generates a longitudinal effective magnetic field $\mathbf{H}_{\mathrm{IFE}}$. This field exists for the laser pulse duration only. The impulsive, non-resonant excitation results in a well-defined initial phase of the spin precession. This process is also effective in fully compensated antiferromagnets  \cite{SatohPRL10,Ivanov14,Satoh15,Bossini17,Tzschaschel17}. Such a simple fundamental excitation is an ideal test case for us. Its well-understood dynamics allow us to demonstrate how to separate thermal electron dynamics from non-thermal spin dynamics and how to obtain the full order-parameter trajectory.

We choose hexagonal \ymo\ as a model system. The material is ferroelectric below $T_\mathrm{C}\approx\SI{1250}{K}$  \cite{Lilienblum15}. The three Mn$^{3+}$ sublattices order antiferromagnetically below the N\'{e}el temperature $T_\mathrm{N}\approx\SI{70}{K}$ and form a quasi-twodimensional triangular lattice with a fully compensated spin structure. The spins in the sublattices point along the local $x$~axes \cite{Fiebig05}. The magnetic space group of the ground state is \gs, which gives rise to a SHG contribution coupling bilinearly to the ferroelectric and the AFM order parameter \cite{Sa2000}. The magnetic structure yields three orthogonal magnon modes, named $X$, $Y$ and $Z$~mode, which differ in the direction of the transiently appearing oscillating uncompensated magnetisation component. The individual modes can be excited optically and probed selectively depending on the setting of pump and probe-light polarisations \cite{Satoh15}. Here we exemplarily focus on the $Z$~mode. 

\begin{figure}
	\includegraphics[width= \columnwidth]{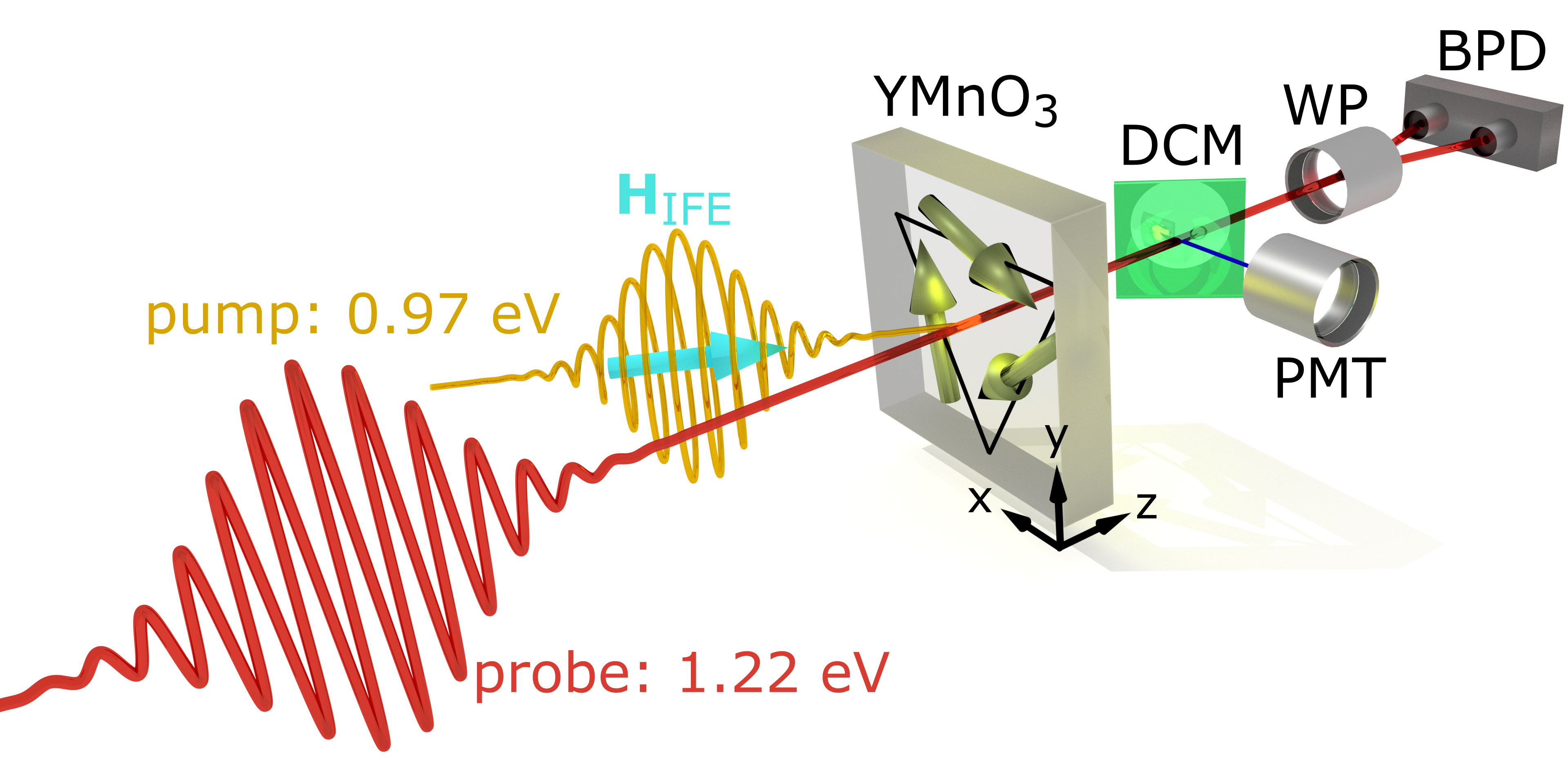}
	\caption{A circularly polarised \SI{130}{fs} laser pulse at \SI{0.97}{eV} induces a spin canting in \ymo\ via the IFE. The subsequent spin precession is probed with a linearly polarised \SI{1.22}{eV} laser pulse via a simultaneous measurement of the Faraday rotation (using a Wollaston prism (WP) and a balanced photodiode (BPD)) and the SHG signal (using a Glan-Taylor prism and a photomultiplier tube (PMT)). DCM: dichroic mirror.}
	\label{fig1}
\end{figure}

Our setup is sketched in Fig.~\ref{fig1}. Via the IFE, a circularly polarised pump pulse cants the spins in the $xy$~plane by an angle $\gamma \mu_0 H_\mathrm{IFE}\tau$, where $\gamma$ is the gyromagnetic ratio, $\mu_0$ is the vacuum permeability and $\tau$ is the duration of the effective magnetic field pulse \cite{LandauLifshitz84}. The transmitted linearly polarised probe pulse is split behind the sample by a dichroic mirror for separate, yet simultaneous, time-resolved detection of the Faraday rotation and of the relative change of the SHG intensity ($\eta = \Delta I_\mathrm{SHG}/I_\mathrm{SHG}$). 

A typical measurement at \SI{10}{K} is shown in Fig.~\ref{fig2}. The Faraday rotation in Fig.~\ref{fig2}a exhibits sine-like behaviour with a frequency of $\Omega/2\pi = \SI{95.2\pm0.3}{GHz}$. The initial phase changes by $\pi$ upon changing the pump helicity, as is expected for a $Z$-mode excitation via the IFE \cite{Satoh15}. Figure~\ref{fig2}b shows that the simultaneously measured SHG response $\eta$ is dominated by two processes. Taking $\mathrm{(\eta_--\eta_+)/2}$ and $\mathrm{(\eta_-+\eta_+)/2}$, with the sign indicating the pump-pulse helicity, we can separate helicity-dependent and helicity-independent contributions, respectively. The latter can be fitted with a single exponential decay. The helicity independence suggests thermalisation processes as a likely origin \cite{Melnikov03}. The helicity-dependent difference reveals cosine-like behaviour with a frequency matching that of the Faraday rotation.  The initial phase reflects the helicity dependence of the excitation mechanism, albeit with a phase shift of $\pi/2$ with respect to the Faraday response. Therefore, the SHG modulation cannot be related to the transient uncompensated magnetisation component. Since the linear transmittance of the sample both at $\omega$ and $2\omega$ does not show any periodic changes either (not shown), we conclude that we observe a modulation of $\mathcal{X}_{ijk}$ induced by dynamic changes of the AFM order parameter $\ell$. There are two fundamentally different mechanisms that can affect $\mathcal{X}_{ijk}$ through $\ell$:

\begin{figure}
	\includegraphics[width= \columnwidth]{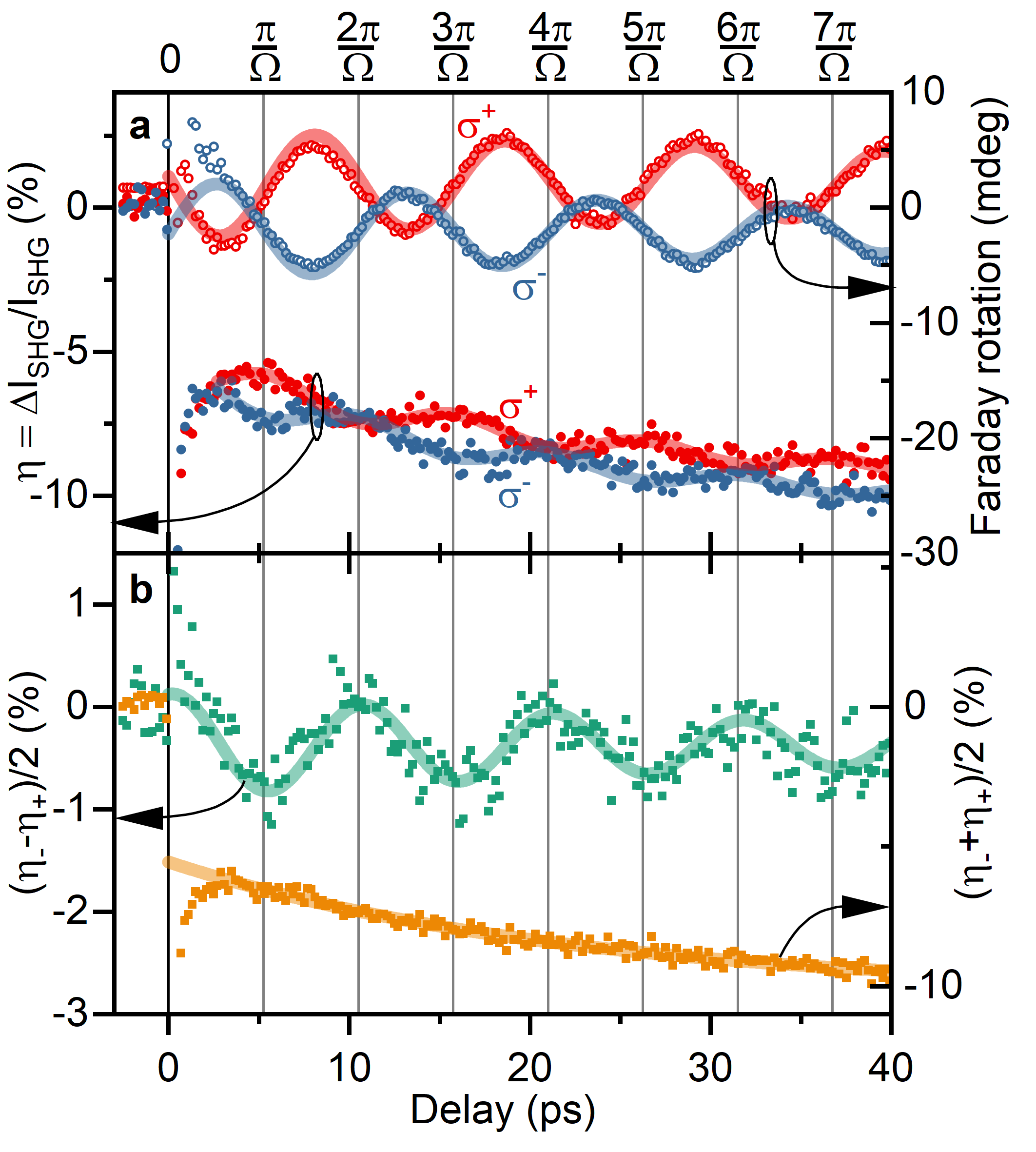}
	\caption{Typical measurement of optically induced spin dynamics by $\sigma^+$-polarised (red) and $\sigma^-$-polarised (blue) pump pulses. \textbf{a} Open circles represent Faraday rotation. Closed circles show the relative change of SHG intensity. Solid lines are damped sine (Faraday) and cosine (SHG) fits. \textbf{b}  Difference (green) and average (orange) of time-resolved SHG changes corroborating helicity-dependent and helicity-independent contributions, respectively. Solid lines represent fits according to a damped cosine (green) and a single exponential decay (orange).}
	\label{fig2}
\end{figure}

\begin{enumerate}[label=(\roman*)]
	\item Symmetry-conserving dynamics involving amplitude modulations of the existing components $\mathcal{X}_{ijk}$. Examples are longitudinal coherent dynamics \cite{Mankowsky17,Sheu18}, changes of the exchange interaction \cite{Melnikov03,Matsubara15} as well as thermally induced quenching dynamics \cite{Satoh07,Sheu16}. 
	\item Symmetry-changing dynamics leading to the appearance of new components $\mathcal{X}_{ijk}$. Besides, e.g., spin-reorientation \cite{Kimel04} and AFM switching processes \cite{Kimel09}, this also includes transversal spin dynamics, where precessional spin motion causes a redistribution among the $\mathcal{X}$ components with periodically arising and vanishing contributions.
\end{enumerate}

\begin{figure}
	\includegraphics[width= \columnwidth]{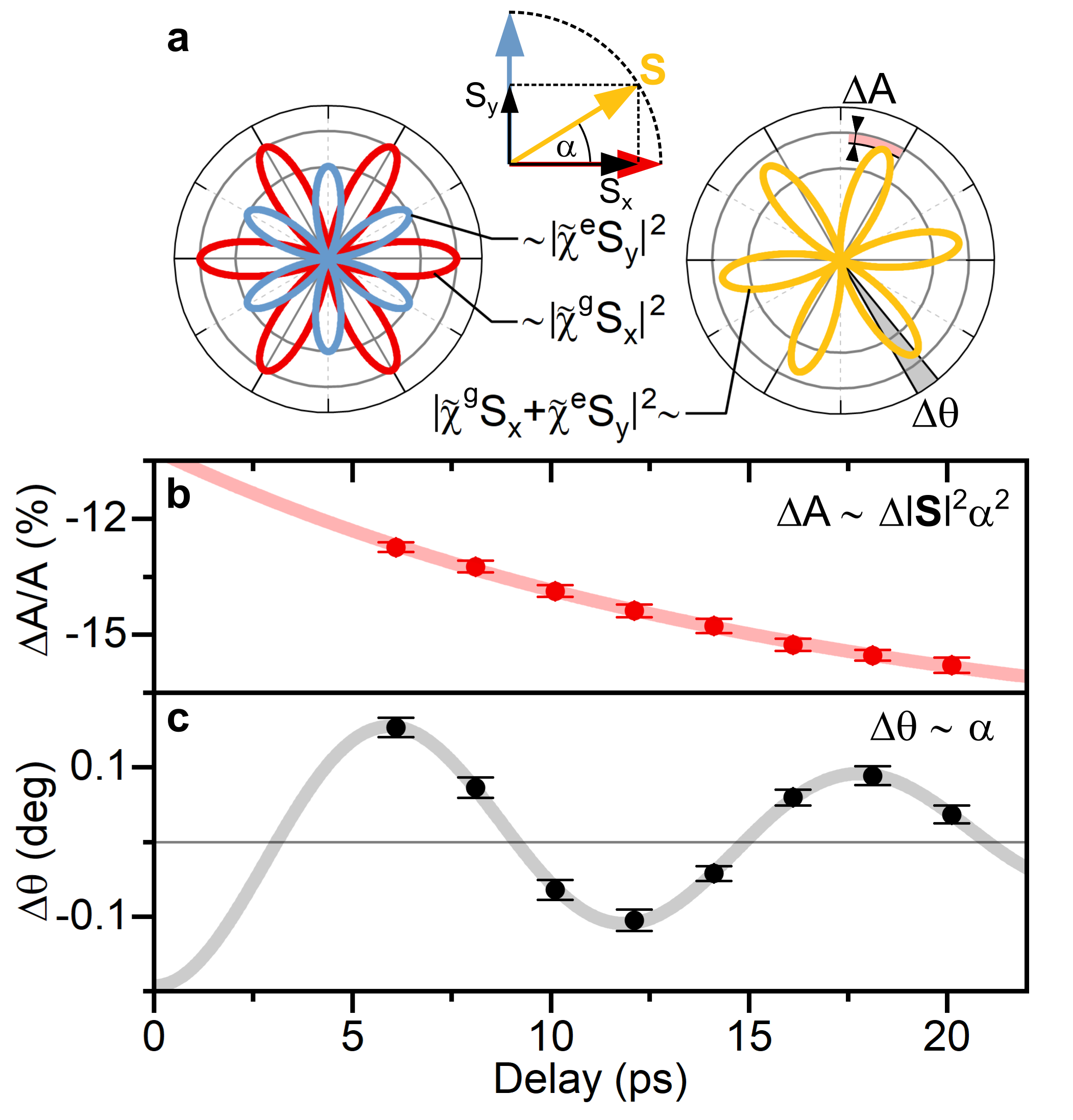}
	\caption{\textbf{a} Red and blue curves are calculated SHG anisotropies according to \gs\ ($\mathbf{S}\parallel\mathbf{x}$) and \es\ ($\mathbf{S}\parallel\mathbf{y}$) symmetry, respectively. A spin canted by an angle $\alpha$ leads to a superposition of $S_x$- and $S_y$-related SHG contributions. Their interference causes a rotation and amplitude change of the SHG anisotropy (yellow curve). \textbf{b} Time dependence of the relative amplitude change. \textbf{c} Time dependent rotation angle of the SHG anisotropy after excitation. Solid lines are fits.}
	\label{fig4}
\end{figure}

We distinguish the two cases by measuring the highly symmetry-sensitive anisotropy of the SHG signal before and after the excitation. Prior to the excitation, the system has the \gs\ symmetry of the ground state with the Mn$^{3+}$ spins oriented along equivalent crystallographic $x$ axis. The corresponding SHG anisotropy in Figs.~\ref{fig4}a (red curve) exhibits six lobes of amplitude $A$ spaced $60^\circ$ each. Its orientation is defined as $\theta = 0^\circ$. Fig.~\ref{fig4}b and \ref{fig4}c show changes of $A$ and $\theta$, respectively after excitation.

The amplitude $A$ exhibits an exponential decrease of the order of 10$\%$. The change of $A$ cannot be caused by changes of the optical properties as the slight changes of the transmittivity at $\omega$ and $2\omega$ are not sufficient to explain the 10$\%$ decrease of $A$. Instead, the isotropic, i.e.\ polarisation independent, behaviour is indicative of an incoherent relaxation. We therefore attribute this decay to longitudinal order-parameter dynamics, i.e.\ a reduction of $\vert\mathbf{S}\vert$ related to the relaxation of photoexcited charge carriers.

Figure~\ref{fig4}c reveals a striking periodic rotation of the SHG anisotropy. The modulation frequency of $\Delta\theta$ matches that of the Faraday rotation, which confirms a magnetic origin. As illustrated in Fig.~\ref{fig4}a, a rotated anisotropy can be decomposed into a linear superposition of two sixfold anisotropies at $\theta = 0^\circ$ (spins along $x$ axis, space group \gs, red curve) and $\theta = 90^\circ$ (spins along $y$ axis, space group \es, blue curve). The superposition corresponds to a coherent deviation of the spins from the $x$ axis towards the $y$ axis and, therefore, to a symmetry reduction to $P6_3^\prime$ or lower. With $\tan\alpha = S_y/S_x$, we can write the dynamic SHG tensor as $\mathcal{X}(t) = \cos\alpha(t)\tilde{\mathcal{X}}^{g}+\sin\alpha(t)\tilde{\mathcal{X}}^{e}$ with time-independent tensors $\tilde{\mathcal{X}}^{g}$ and $\tilde{\mathcal{X}}^{e}$. Thus, complementary to the Faraday rotation, which is sensitive to the uncompensated $z$~component of the spin precession, the rotation of the SHG anisotropy reflects the transversal dynamics of the compensated spin component in the $xy$~plane. Furthermore, according to the Curie principle, the isotropic effect of a thermal excitation cannot reduce the symmetry of $\mathcal{X}$. Thus, the SHG anisotropy rotation is a pristine measure of the non-thermal spin-dynamics. Separating thermal and non-thermal dynamics is therefore possible.

We can use the SHG measurements to quantify the spin deflection angle $\alpha$ according to $3\Delta\theta \approx \rho\alpha$, where $\rho = \vert\tilde{\mathcal{X}}^{e}\vert/\vert\tilde{\mathcal{X}}^{g}\vert = 0.6$ is the amplitude ratio of the real-valued susceptibilities $\tilde{\mathcal{X}}_{ijk}$ \cite{Fiebig00b,Iizuka01,Lottermoser02}. Note that $\rho$ changes by $\pm0.3$ within the probe laser linewidth, which introduces a systematic error of that order. From Fig.~\ref{fig4}a, we extrapolate $\Delta\theta(0) = 0.198^\circ\pm0.007^\circ$ and, hence, $\alpha(0) = 1.0^\circ$ with a statistical and systematic uncertainty of $0.035^\circ$ and $0.5^\circ$, respectively. This in turn allows us to quantify the effective magnetic field of the inverse Faraday effect as $\alpha(0) = \gamma \mu_0 H_\mathrm{IFE} \tau$, where $1^\circ$ corresponds to $\mu_0 H_\mathrm{IFE} \approx \SI{760}{mT}$ for $\tau = \SI{130}{fs}$ (assuming the free electron value for $\gamma$). Note that the optically induced spin canting by non-resonant excitation via the IFE is comparable to resonant excitation via the magnetic field component of a strong THz pulse \cite{Baierl16}.

In addition to the quantification of the basal-plane spin-canting angle, we can use the time-resolved Faraday rotation to estimate the maximum out-of-plane spin canting angle as $\mathrm{1.4\,mdeg}$. Such a highly anisotropic spin precession is a general phenomenon in antiferromagnets and highlights the dominance of the exchange interaction over the weak magnetic in-plane anisotropy \cite{Toulouse14,Tzschaschel17}. 

\begin{figure}
	\includegraphics[width= \columnwidth]{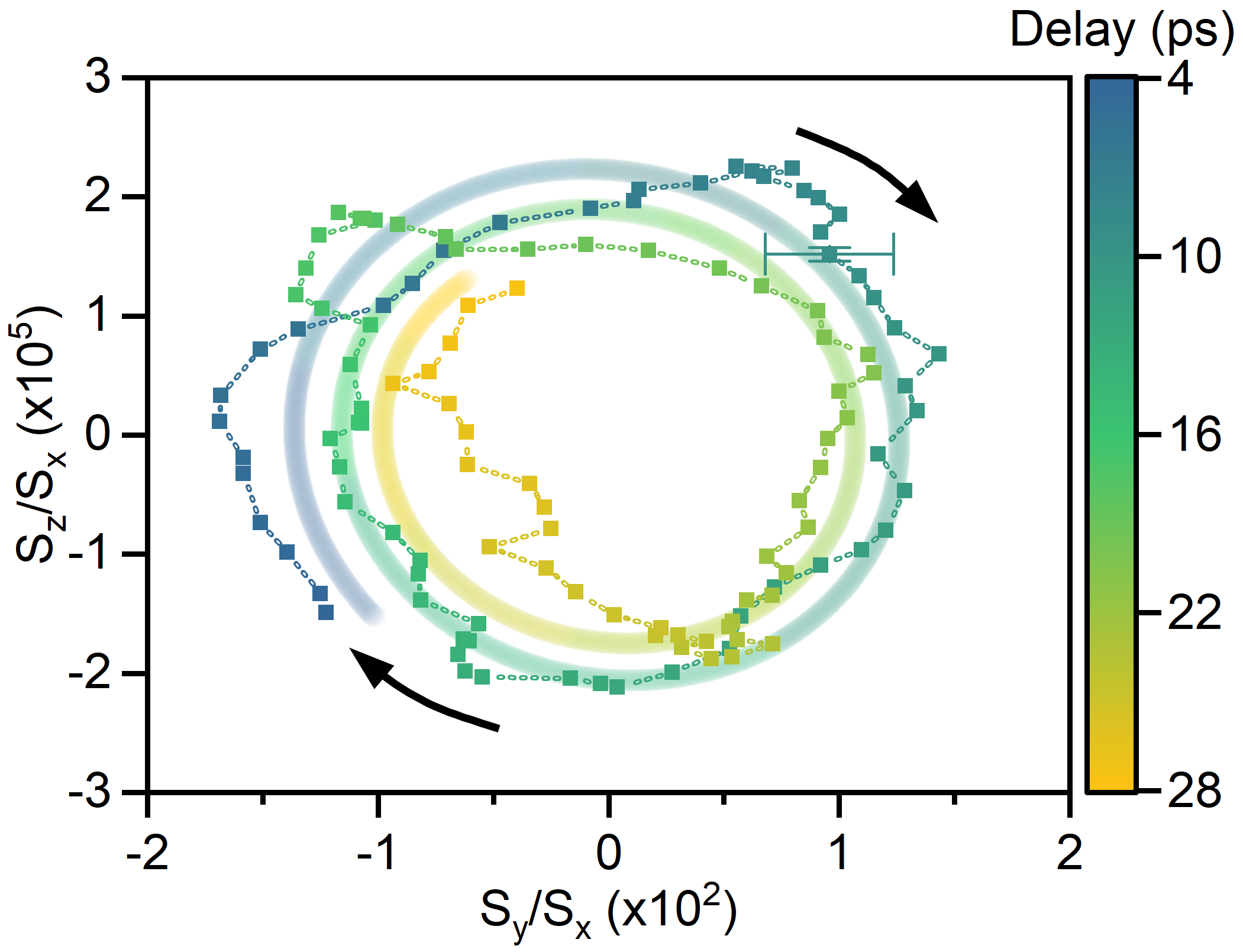}
	\caption{Time-dependent amplitude of the spin motion in the $yz$ plane obtained by combining floating time averages of SHG and Faraday rotation measurements. The solid line is a fit of a damped elliptical spin precession. For clarity, only two precession periods are shown.}
	\label{fig5}
\end{figure}

After relating the basal-plane spin-canting angle $\alpha$ and the total sublattice magnetisation $\vert\mathbf{S}\vert$ to the SHG anisotropy orientation and amplitude, respectively, we understand the measurement shown in Fig.~\ref{fig2} as follows: The observed TR-SHG response is a superposition of longitudinal and transversal spin dynamics and therefore a combination of scenarios (i) and (ii). The time-dependence of $\alpha$ and hence $S_y$ is reflected by the periodic SHG modulation. We combine this with the $S_z$-sensitive Faraday rotation to obtain the full order-parameter motion. By plotting the transversal spin components in Fig.~\ref{fig5}, one can clearly recognise the magnonic damped elliptical motion (solid line).

We thus present a proof-of-concept experiment, where we track the full three-dimensional motion of an antiferromagnetic order parameter. In our model compound, \ymo, SHG and Faraday rotation are combined to obtain the trajectory of an optically induced coherent spin precession. While the out-of-plane spin canting induces an uncompensated magnetic moment that causes a Faraday rotation, the basal-plane canting and the total sublattice magnetisation are reflected in the SHG measurements. We quantify the optically induced canting as approximately $1^\circ$, corresponding to an effective magnetic field of the inverse Faraday effect of \SI{760}{mT}. Key to this analysis is the transient symmetry reduction during coherent excitations. This directly affects the highly symmetry sensitive SHG anisotropy and allows us to separate non-thermal spin dynamics from thermal electron dynamics. We emphasise though that our approach is not at all limited to harmonic small-amplitude dynamics. Instead, we exploit a fundamental excitation to showcase the insights provided by a time-resolved symmetry analysis, while at the same time establishing a general approach for accessing order-parameter dynamics at sub-picosecond timescales. Tracking the full order-parameter motion instead of just one component is indispensable for understanding the highly complex dynamics occurring during ultrafast switching, spin-reorientation and other non-equilibrium phenomena.

The authors thank Morgan Trassin and Mads Weber for valuable discussions. T.S. was supported by Japan Society for the Promotion of Science (JSPS) KAKENHI (JP26103004) and JSPS Core-to-Core Program (A. Advanced Research Networks). C.T. and M.F. acknowledge support from the SNSF project 200021/147080
and by FAST, a division of the SNSF NCCR MUST. M.F. thanks ETH Zurich and CEMS at RIKEN for support of his research sabbatical.

\end{document}